\documentclass[usenatbib]{mn2e}
\usepackage{amsmath,amssymb}
\usepackage[T1]{fontenc}
\usepackage{aecompl}
\usepackage{graphicx}
\usepackage{color}
\title[{\color{black} Radiation fields by} IA SPs with binaries as ionizing sources of H II regions]
{{\color{black} Radiation fields by intermediate-age stellar populations with} binaries as ionizing sources of H II regions}

\author[F. Zhang, L. Li, L. Cheng, L. Wang, X. Kang, Y. Zhuang, Z. Han]
{
F. Zhang\thanks{E-mail: zhangfh@ynao.ac.cn; zhang\_fh@hotmail.com}$^{1,2}$, L. Li$^{1,2}$, L. Cheng$^{1,2,3}$, L. Wang$^{1,2,3}$, X. Kang$^{1,2,3}$, Y. Zhuang$^{1,2,3}$, Z. Han$^{1,2}$\\
$^1$Yunnan Observatories, Chinese Academy of Sciences, Kunming, 650011, China \\
$^2$Key Laboratory for the Structure and Evolution of Celestial Objects, Chinese Academy of Sciences, Kunming, 650011, China \\
$^3$University of Chinese Academy of Sciences, Beijing, 100049, China}

\begin{document}
\date{\today}
\pagerange{\pageref{firstpage}--\pageref{lastpage}}
\pubyear{2011}
\maketitle
\label{firstpage}

\begin{abstract}
{\color{black} Radiation fields emitted by} O, B-type stars or young stellar populations (SPs) are generally considered as {\color{black} significant} central ionizing sources (CISs) of classic H II regions. 
{\color{black} In our previous studies, we show that the inclusion of binary interactions in stellar population synthesis models can significantly increase the ultraviolet spectrum hardness and the number of ionizing photons of intermediate-age (IA, 7$\la$log($t$/yr)$\la$8) SPs.} 
{\color{black} In this work, we present photoionization models of H II regions ionized by radiation fields emitted by IA SPs, and}
show that radiation fields of IA SPs, including binary systems, {\color{black} are in theory possible candidates of significant} CISs of classic H II regions.

When {\color{black} radiation fields of} IA SPs comprising binary systems are used as the CISs of classic H II regions, {\color{black} the theoretical strengths of a number of lines} (such as {\color{black} [O III]}$\lambda$4959$'$, [S II]$\lambda$6716$'$, etc.), weaker than observations, are raised; the border /selection-criterion lines, between star-forming galaxies and AGNs in the diagnostic diagrams (for example, [N II]$\lambda$6583/H$\alpha$$\sim$[O III]$\lambda$5007/H$\beta$), move into the region occupied originally by AGNs; and He II $\lambda$1640 line, observed in Lyman break and high-redshift gravitationally lensed galaxies, also can be produced.
\end{abstract}


\section{Introduction}
\label{Sect:}
An H II region is a large low-density cloud of partially ionized gas and originated from a giant molecular cloud. H II regions always associate with regions of  recent star formation, can ionize surrounding gas and produce radio emission and various emission lines, so they relate closely to the determinations of star formation rate and chemical compositions of galaxies.

In general, it is thought that the {\color{black} significant} central ionizing sources ({\color{magenta} CISs}) of classic H II regions are {\color{black} radiation fields emitted by} O, B-type stars or clusters of such stars (i.e., young stellar populations, {\color{magenta} SPs}) and that the intermediate-age ({\color{magenta}IA}) and old SPs have no enough hard extreme ultraviolet ({\color{magenta} EUV}) spectra to ionize neutral hydrogen.
For example, in the photoionization models of \citet[hereafter {\color{magenta} Dop00}]{dop00} and \citet[hereafter {\color{magenta} Kew01}]{kew01}, the CISs of H II regions were {\color{black} the radiation fields emitted by} SPs from ages of 0 to 6 and of 0 to 8\,Myr, corresponding to the evolutionary population synthesis ({\color{magenta}EPS}) models of {\sl P\'EGASE} \citep[using Padova tracks]{fio97,fio99} and {\sl Starburst99} \citep[Geneva tracks]{lei99}, respectively.
They thought that there is no further evolution for SPs' EUV spectra and a balance between star birth and star death is set up for all stellar masses contributing significantly to the SPs' EUV spectra.

In fact, IA and old SPs are very important for classic H II regions. Starburst would last $\sim$$10^8$\,yr (comparable to the typical dynamical timescale) and old SPs will dominate the spectrum because young SPs have dispersed into field during this period \citep[][hereafter {\color{magenta} G08}]{gro08}. 
\citet{kan14} have ever used the $t$=2\,Gyr old SP's spectra from the EPS models of \citet{bru03} to investigate the effect of local photoionizing radiation on gas cooling rate. The cooling rate of halo gas is critical to determine how much fuel is available to form stars in galaxies. 
Their work based on the facts that the spectra of SPs older than 200\,Myr are hard due to the accumulation of post asymptotic giant branch ({\color{magenta} AGB}) stars and that the spectrum shape remains fairly constant from 200\,Myr to 13 Gyr because low-mass stars evolve within a narrow temperature range all the way from main sequence ({\color{magenta} MS}) to AGB phase. 
Moreover, G08 also investigated the effect of old SPs on their photoionization models.

In previous studies, in which SPs{\color{black}'s radiation fields are} the CISs of classic H II regions, a SP is solely composed of single stars and has no {\color{black} binaries. Binaries} are a reason of hardening SPs' spectra, thus varying the properties of the CISs and H II regions. 
\citet{kew13} have mentioned a scenario that binary interactions ({\color{magenta} BIs}) harden the SP's spectra: BIs make the stellar systems hotter and more luminous by spinning-up the rotation of the companion star and producing mixing effect \citep{jia14}. 
In fact, BIs can harden and raise the SPs' EUV spectra by other processes (such as merge) via producing very hot and luminous stars.

In this study, we will use Yunnan-II EPS models  \citep[hereafter {\color{magenta} Z04, Z05}]{zha04,zha05a}, in which various BIs are considered, 
{\color{black} to show that  radiation fields of IA SPs with BIs are in theory possible candidates of significant CISs of classic H II regions.}
%
Moreover, we will show
that He II $\lambda$1640 feature (He II feature can be produced by shocks, X-ray binary evolution and WR/massive stars, \citealt{bri08}) can be produced, some theoretical line strengths weaker than observations will be increased and the border lines between star-forming galaxies and AGNs in the diagnostic diagram will move into the region occupied originally by AGNs.

The outline of the paper is as follows. In Section 2 we describe the used photoionization, EPS models and parameter space. In section 3 we present 
{\color{black} the results.}
In Section 4 we present analyses and discussions. Finally we present a summary and conclusions in Section 5.

\section{Models}
\label{Sect:}
\begin{table}
\tiny
\centering
\caption{Part of model parameters used in this study. 'Case', in row one, denotes the condition that BIs are taken into account. Top and bottom sub-parts are for EPS models and nebulae.}
\begin{tabular}{lrr rrr rr}
\hline
\hline
                            & \multicolumn{5}{c}{EPS models}  \\
Case                  &  without     &    with   &          &           &          &          &           \\
log($t$/yr)         &  6.3    &   6.5    & 6.6    &  ...      & 10.0 & 10.1 & 10.18 \\
$Z$                    & 0.0001& 0.0003 & 0.001& 0.004& 0.01& 0.02 & 0.03   \\
\hline
                           & \multicolumn{5}{c}{nebulae}  \\
log$U$              &  $-4.$  &    $-3.$  & $-2.$&  $-1.$&    0.  &         &             \\
$n_{\rm H}$(cm$^{-3}$)                & 10.  & 100.& 350.&  &    &    &              \\
\hline
\end{tabular}
\label{Tab:space}
\end{table}

\begin{table*}
\centering
\caption{The logarithmic number-abundance ratios of the elements with respect to hydrogen [log($n_{X} \over n_{\rm H}$)$_{Z'_{\odot}}$] and logarithmic element-depletion factors [log($D_{X}$)] at 'solar metallicity' $Z'_{\odot}$=0.016.}
\begin{tabular}{lrr rrr rrr rrr rrr rrr r}
\hline 
 $X$  &  \ \   H\,\,\, &     He\,\,\,\; &    C\,\,\,\; &     N\,\,\,\; &     O\,\,\,\;  &    Ne\,\,\,\; &    Na\,\,\, &     Mg\,\,\, &    Al\,\,\,\; &    Si\ \ \ \; &    S\ \ \ &    Cl\ \ \ &    Ar\ \ \ &    Ca\ \ \ &    Fe\ \ \ &    Ni\ \\
%
\hline
log(${n_{X} \over n_{\rm H}}$)$_{Z'_{\odot}}$ &\multicolumn{18}{l}{$0.00$, $-1.01$, $-3.59$, $-4.22$, $-3.34$, $-3.91$,$-5.75$, $-4.47$, $-5.61$, $-4.49$, $ -4.79$, $-6.40$, $-5.20$, $-5.64$, $-4.55$, $-5.68$} \\
log($D_{X}$)         & \multicolumn{18}{l}{$0.00$, $\  \ 0.00$, $-0.15$, $-0.23$ ,$-0.21$, $\ \ 0.00$, $ -1.00$, $-1.08$, $-1.39$, $-0.81$, $-0.08$, $-1.00$, $  \ \ 0.00$, $-2.52$, $ -1.31$, $-2.00$} \\
\hline
\end{tabular}
\label{Tab:x-d}
\end{table*}

In this work, we employ {\sl MAPPINGS IIIq} version photoionization code
, which is used to calculate {\color{black} the fluxes} of emission lines in H II regions, and Yunnan-II EPS models (Z04, Z05),  {\color{black} from which the SPs' radiation fields are obtained as} the CISs of H II regions in {\sl MAPPINGS IIIq} code. 
In the following, we will briefly describe {\sl MAPPINGS} code and Yunnan-II models in Sections~\ref{Subsect:pho} and \ref{Subsect:eps}, finally we will present parameter space {\color{black} in Section 2.3}.

\subsection{Photoionization code: {\sl MAPPINGS}}
\label{Subsect:pho}
%
{\sl MAPPINGS} code is built by \citet{dop76} and developed by Luc Binette (Mexican version),  \citet{sut93}, \citet[][]{gro04}, G08, and so on.
In this work, the parameters employed in {\sl MAPPINGS IIIq} code are as follows.
%
%
(i) We use  {\color{black} the radiation fields of SPs from} Yunnan-II EPS models with and without BIs as the CISs of H II regions.  The chosen SPs' age and metallicity are within 6.3$\le$log($t$/yr)$\le$10.18 and 0.0001$\le$$Z$$\le$0.03 (Table~\ref{Tab:space}).
(ii)  Plane-parallel geometry is used and dimensionless ionization parameter log$U$, used to define ionizing radius, is from $-4$ to 0 (Table~\ref{Tab:space}).
(iii) Isochoric structure is used and hydrogen density $n_{\rm H}$ ranges from 10 to 350\,cm$^{-3}$ (Table~\ref{Tab:space}).
(iv) The default sets of $'$solar metallicity$'$ ($Z'_{\odot}$=0.016, \citealt{asp05}) element abundances $({n_{X} /n_{\rm H}})_{Z'_\odot}$ and depletion factors $D_{X}$ are used (Table~\ref{Tab:x-d}, same to \citealt[][hereafter {\color{magenta} D06}]{dop06a}).
At other metallicities, $n_{X}/n_{\rm H}$ scales with $(n_{X}/n_{\rm H})_{\rm Z'_\odot}$ and metallicity with several exceptions ({\color{black} He}, C and N elements, follows D06){\color{black}, $D_{X}$ does not vary with metallicity}.
%
(vi) Three kinds of dust components are included: graphites, silicates and PAHs (G08). For the two formers, the size distribution of ${{\rm d} n_X \over {\rm d}a}$=${\rm k}\,a^{-\beta}\,{e^{-(a/a_{\rm min})^{-3}} \over 1+e^{-(a/a_{\rm max})^{3}}}$ ($\beta$=3.3, $a_{\rm min}$=40\,\AA, $a_{\rm max}$=1600\,\AA and k=const) is used and {\color{black} the density of themselves} is 1.8 and 3.5\,g/cm$^3$.
%
{\color{black} For PAHs, we assume that graphites are not destroyed with PAHs, PAHs emit in the mid-infrared band when $Q_2$ (a far-UV analogy of $U$)$<$1000 \citep{dop05}, the PAH-to-carbon dust ratio is 0.3 and the fraction of carbon dust depletion in PAHs is 0.05.}
(vii) At last, 
{\color{black} we choose the set of input physics including radiation pressure and dust},
use the geometrical dilution fraction of 0.5 and use a final total hydrogen column depth of 10$^{22}$\,cm$^{-2}$.

\subsection{EPS models: Yunnan-II}
\label{Subsect:eps}
Yunnan-II models are built since Z04 and Z05 and {\color{black} are the EPS models for instantaneous burst SPs (i.e., assemblies of chemically homogeneous and coeval stars) without (i.e., simple SPs) and with BIs. For the second set of models}, various BIs are considered {\color{black} (mass transfer, mass accretion, common-envelope evolution, collisions, supernova kicks, tidal evolution, and all angular momentum loss mechanisms)}.
{\color{black} In Z04 and Z05 models, each binary in a SP satisfies the given initial primary-mass, mass-ratio $q$, separation $a$ and eccentricity $e$ distributions, evolutionary parameters (gravity, temperature, etc.) are obtained by evolution algorithm, observables (spectrum, colours, etc.) are transformed from evolutionary parameters via stellar spectral library, at last the SPs' properties are from the integral between observables and weight given by initial distributions.}
The ages and metallicities of SPs cover the ranges log($t$/yr)=5.0-10.18 and $Z$=0.0001-0.03. 
%

Yunnan-II models include several sets of results, the differences among them mainly lie in the choices of initial distributions {\color{black} for stars in a SP}, stellar spectral library, and so on.
The models, employed in this work, use BaSeL stellar spectral library of \citet{lej97}, binary star evolution algorithm of \citet{hur02}, the initial mass function of \citet{mil79} for primary star, uniform $q$ distribution, the combination of power-law and const $a$ distributions at close and far separations and uniform $e$ distribution. The separation distribution implies that  $\sim$50 per cent (a typical value for the Galaxy) of stellar systems are binary systems with orbital periods less than 100\,yr.

\subsection{Parameter space}
\label{Subsect:space}
In this work, we assume that the metallicities of CISs (i.e., SPs) are same as those of nebular gas (same as \citealt[][hereafter {\color{magenta} Sta06}]{sta06}) and our calculations are only limited to these SPs' metallicities. 
%
%
{\color{black} The element abundances at SP's metallicities $(n_{X}/n_{\rm H})_Z$ are obtained 
{\color{black} as described in Section 2.1.}

In total, we calculate 8190 models in this study (Table~\ref{Tab:space}). {\color{black} Two sets of SP's radiation fields, from Yunnan-II EPS models} with and without BIs, are used as the CISs. For each set of EPS models, we choose 39 age values [6.3$\le$log($t$/yr)$\le$10.18 in steps of 0.1/0.08 or 0.2] and all metallicity values ($Z$=0.0001, 0.0003, 0.001, 0.004,  0.01, 0.02 and 0.03). 
Given a {\color{black} radiation field}, five log$U$ values ($-4.0, -3.0, -2.0, -1.0$ and 0.) and three $n_{\rm H}$ values (10, 100 and 350\,cm$^{-3}$) are used.

\section{Results}
\label{Sect:}
\begin{table}
\tiny
\caption{List of emission lines (top-), constituent lines (middle-), line strength ratios and grids (bottom-subparts) analyzed in this work.}
\begin{tabular}{lllll}
\hline
\multicolumn{5}{c}{\bf emission lines} \\
 $\rm [O \ II]        \lambda$3727      & [O \ II]$\lambda$3729     & [NeII]$\lambda$3869                 & [S \,II]$\lambda$4072   & [OIII]$\lambda$4363   \\
 $\rm [H \ \,\beta]\lambda$4868     & [O\ III]$\lambda$4959     & [OIII]$\lambda$5007                  & [N II]$\lambda$5755     & [O \ I]$\lambda$6300  \\
 $\rm [S\,III]         \lambda$6312     & [N \ II]$\lambda$6548     & [H\ \,$\alpha$]$\lambda$6563  & [N II]$\lambda$6583    & [S \,II]$\lambda$6716  \\
 $\rm [S\,\ II]        \lambda$6731     & [ArIII]$\lambda$7135      & [O II]$\lambda$7320$^{\rm \color{magenta}b1}$   
                                                                                                         & [O II]$\lambda$7331$^{\rm \color{magenta}b2}$                         & [S\,III]$\lambda$9069  \\
$\rm [S\,III]\lambda$9532 & & & & \\
\end{tabular}

\begin{tabular}{lllll}
\multicolumn{5}{c}{\bf constituent lines} \\
\multicolumn{5}{l}{ 
 $\rm [O  \ II]  \lambda3727'^{\color{magenta} c1}$  
 $\rm [OIII]     \lambda4959'^{\color{magenta} c2}$  
 $\rm [N II]     \lambda6548'^{\color{magenta} c3}$  
 $\rm [S\,\ II] \lambda6716'^{\color{magenta} c4}$ 
 $\rm [O II]    \lambda7320'^{\color{magenta} c5}$ }\\
\multicolumn{5}{l}{ 
 $\rm  [S\,III] \lambda9069'^{\color{magenta} c6} $   
 $\rm  R_{23}   $$               ^{\color{magenta} c7} \qquad \quad \ \; $   
 $\rm  S_{23}   $$               ^{\color{magenta} c8} \qquad \quad \ $
 $\rm  N_2        $$              ^{\color{magenta} c9} \qquad \quad \ \; \ $  
 $\rm  O_3N_2$$              ^{\color{magenta} c10}$ }    \\

\multicolumn{4}{c}{\bf line strength ratios and grids} \\
\multicolumn{4}{l}{ 
$\rm {[OIII]\lambda5007\ \ \                           \over [\rm O\ II]\lambda3727'^{\color{cyan}c1}}^{\color{magenta} d1}$      
$\rm {[SIII]\lambda9069'^{\color{cyan}c6}  \over [\rm S\ II]\lambda6716'^{\color{cyan}c4}}^{\color{magenta} d1,2}$  
$\rm {[S II]\lambda6716                                 \over [\rm S II]\lambda6731                              }^{\color{magenta} d3}$     \qquad \quad
$\rm {[O II]\lambda3727                                 \over [\rm O II]\lambda7325                              }^{\color{magenta} d4}$  }   \\

\multicolumn{4}{l}{ 
$\rm {[N \ II]\lambda6548'^{\color{cyan}c3}   \over [\rm N \ II]\lambda5755\ \ \ \                       }$$^{\rm \color{magenta} d4}$    
$\rm {[SIII]  \lambda9069'^{\color{cyan}c6}    \over [\rm SIII] \lambda6312\ \ \ \                        }$$^{\rm \color{magenta} d4}$     \;\;
$\rm {[S II]  \lambda6716'^{\color{cyan}c4}    \over [\rm S II] \lambda4072\ \ \ \                        }$$^{\rm \color{magenta} d4}$     \quad \;
$\rm {[O II]  \lambda3727\ \ \ \                           \over [\rm O II] \lambda7320'^{\color{cyan}c5}}$$^{\rm \color{magenta} d5}$ }    \\

\multicolumn{4}{l}{ 
$\rm {[OIII]\lambda4959'^{\color{cyan}c2}   \over [\rm OIII]  \lambda4363\ \ \ \                        }$$^{\rm \color{magenta} d6}$     
$\rm {[N \ II]\lambda6583\ \ \ \                        \over [\rm O \ II]\lambda3727'^{\color{cyan}c1}}$$^{\rm \color{magenta} d7}$    \;
$\rm {[N II]\lambda6583\ \ \ \                          \over [\rm S II]  \lambda6716'^{\color{cyan}c4}}$$^{\rm \color{magenta} d7,8}$\quad
$\rm {[OIII]\lambda4959'                                 \over R_{23}*H\beta}^{\rm \color{magenta}d9}$                                                                                                 } \\

\multicolumn{4}{l}{ 
$\rm {R_{23}}^{\rm \color{cyan}c7,\color{magenta}d7,10}$  \qquad \;
$\rm {S_{23}}^{\rm \color{cyan}c8,\color{magenta}d7}$        \qquad \qquad \;
$\rm {N_2}$$^{\rm \color{cyan}c9,\color{magenta}d7,11}$   \qquad \qquad
$\rm {O_3N_2}$$^{\rm \color{cyan}c10,\color{magenta}d7,11}$ }\\
 
\multicolumn{4}{l}{ 
$\rm {[O\, II]\lambda}7320'^{\color{cyan}c5,}$$^{\rm \color{magenta} d12}$ 
$\rm {[O\, II]\lambda}3727 $$^{\rm \color{magenta} d12} \qquad $ 
$\rm {\color{black} [O III]\lambda4959'^{\color{cyan}c2,}}$$^{\rm \color{magenta} d13}$ }\\
\hline
\end{tabular}\\
%
%
\begin{tabular}{llll}
\multicolumn{2}{l}{$^{\rm \color{magenta} b1}$$\lambda$7319.0 \& 7320.0 two lines; }& 
\multicolumn{2}{l}{$^{\rm \color{magenta} b2}$$\lambda$7330.1 \& 7330.7 two lines.} \\
\multicolumn{4}{l}{\color{magenta} Note, beginning with 'c', gives the definition.}\\
\multicolumn{2}{l}{$^{\rm \color{magenta} c1} \rm {[O  II]\lambda3727'}$=[O II]$\lambda$3727+3729; } & 
\multicolumn{2}{l}{$^{\rm \color{magenta} c2} \rm {[OIII]\lambda4959'}$=[OIII]$\lambda$4959+5007; } \\
\multicolumn{2}{l}{$^{\rm \color{magenta} c3} \rm {[N II]\lambda6548'}$=[N II]$\lambda$6548+6583; }&
\multicolumn{2}{l}{$^{\rm \color{magenta} c4} \rm {[S\ \,II]\lambda6716'}$=[S II]$\lambda$6716+6731; } \\
\multicolumn{2}{l}{$^{\rm \color{magenta} c5} \rm {[O II]\lambda7320'}$=[O II]$\lambda$7320+7331; } &
\multicolumn{2}{l}{$^{\rm \color{magenta} c6} \rm {[S\,III]\lambda9069'}$=[S\,III]$\lambda$9069+9532;}  \\
\multicolumn{2}{l}{$^{\rm \color{magenta} c7} \rm R_{23}$=$\rm [OII]\lambda3727'+[OIII]\lambda4959' \over H\beta$;} &
\multicolumn{2}{l}{$^{\rm \color{magenta} c8} \rm S_{23}$=$\rm [SII]\lambda6716'+[SIII]\lambda9069' \over H\beta$;} \\
\multicolumn{2}{l}{$^{\rm \color{magenta} c9} \rm {N_2 }$=${[\rm N II]\lambda6583}/{\rm H\alpha}$; }& 
\multicolumn{2}{l}{$^{\rm \color{magenta} c10} \rm {O_3N_2}$=${[\rm OIII]\lambda5007/H\beta \over [\rm N \ II]\lambda6583/H\alpha }$; }\\
\multicolumn{4}{l}{\color{magenta} Note, beginning with 'd', denotes which parameter is sensitive to. }\\
$^{\rm \color{magenta} d1}$$U$ (KD02, M); &
$^{\rm \color{magenta} d2}$better; &
$^{\rm \color{magenta} d3}$$n_{\rm e}$ (I06, $f$); & 
$^{\rm \color{magenta} d4}$$T_{\rm e}$ (B05); \\
$^{\rm \color{magenta} d5}$$T_{\rm e,O2}$\,(I06); & 
\multicolumn{2}{l}{$^{\rm \color{magenta} d6}$$T_{\rm e,O3}$\,(I06, $f$($n_{\rm e}$));} &
$^{\rm \color{magenta} d7}$$Z$ (KD02, M); \\
$^{\rm \color{magenta} d8}$$Z$ (D06, M); &
$^{\rm \color{magenta} d9}$$Z$ (P05, O);  &
$^{\rm \color{magenta} d10}$$Z$ (P01, O);  &
$^{\rm \color{magenta} d11}$$Z$ (PP04, O); \\
\multicolumn{2}{l}{$^{\rm \color{magenta} d12}$$ {\rm O^{2+}/H^+}$ (I06, $f(T_{\rm e}$, $n_{\rm e}$));}   &
\multicolumn{2}{l}{$^{\rm \color{magenta} d13}$$ {\rm O^+/H^+}$ (I06, $f(T_{\rm e}$)).}   \\

\multicolumn{4}{l}{\color{magenta} In notes 'd',  the meanings of the abbreviations in parenthesis are as follows.}\\
\multicolumn{2}{l}{{\color{magenta} I06:   \ \ }\citet{izo06};} &
\multicolumn{2}{l}{{\color{magenta} KD02:}\citet{kew02};}\\
\multicolumn{2}{l}{{\color{magenta} P01:  \ }\citet{pil01};} &
\multicolumn{2}{l}{{\color{magenta} P05:  \ }\citet{pil05};}\\
\multicolumn{2}{l}{{\color{magenta} PP04:}\citet{pet04};} &
\multicolumn{2}{l}{{\color{magenta} M: \ \ \,} photoionization models;} \\
\multicolumn{2}{l}{{\color{magenta} O: \ \ \,} observation; }  &
\multicolumn{2}{l}{{\color{magenta} $f$: \ \ \ } function.}  \\

\multicolumn{4}{l}{{\color{magenta} Other notes:} O$_3$N$_2$ of d7 is multiplied by ${\rm H_\beta / H_\alpha}$.}\\
\multicolumn{4}{l}{}  \\

\end{tabular}

\label{Tab:line-list}
\end{table}

\begin{figure}
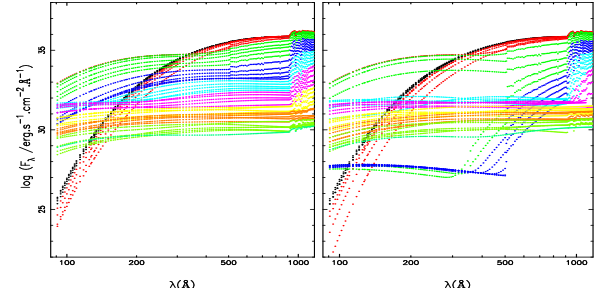

\leftline{
\includegraphics[bb= 51  63 531 699,clip,height=3.8cm,width=4.0cm,angle=0]{ISEDb.ps} 
\includegraphics[bb=109 63 531 699,clip,height=3.8cm,width=3.6cm,angle=0]{ISEDs.ps}
}
\caption{Spectrum evolution (91$\le$$\lambda$/{\rm \AA}$\la$1100) for SPs with (left panel) and without (right panel) BIs at $Z$=0.02. In each panel, 45 values of age are included, each color comprises 5 age values, and the age steps are 0.2 and 0.1/0.08 when log($t$/yr)$\le$6.5 and $>$6.5. 
In the order of SP's age, they are  black [log($t$/yr)=5.1-5.9], red (6.1-6.7), green (6.8-7.2), blue (7.3-7.7),  cyan (7.8-8.2), magenta (8.3-8.7), yellow (8.8-9.2), orange (9.3-9.7) and light-green (9.8-10.18)  lines.}
\label{Fig:ised-bi}
\end{figure}

\begin{figure}
\leftline{
\includegraphics[bb= 50  57 531 698,clip,height=3.8cm,width=4.0cm,angle=0]{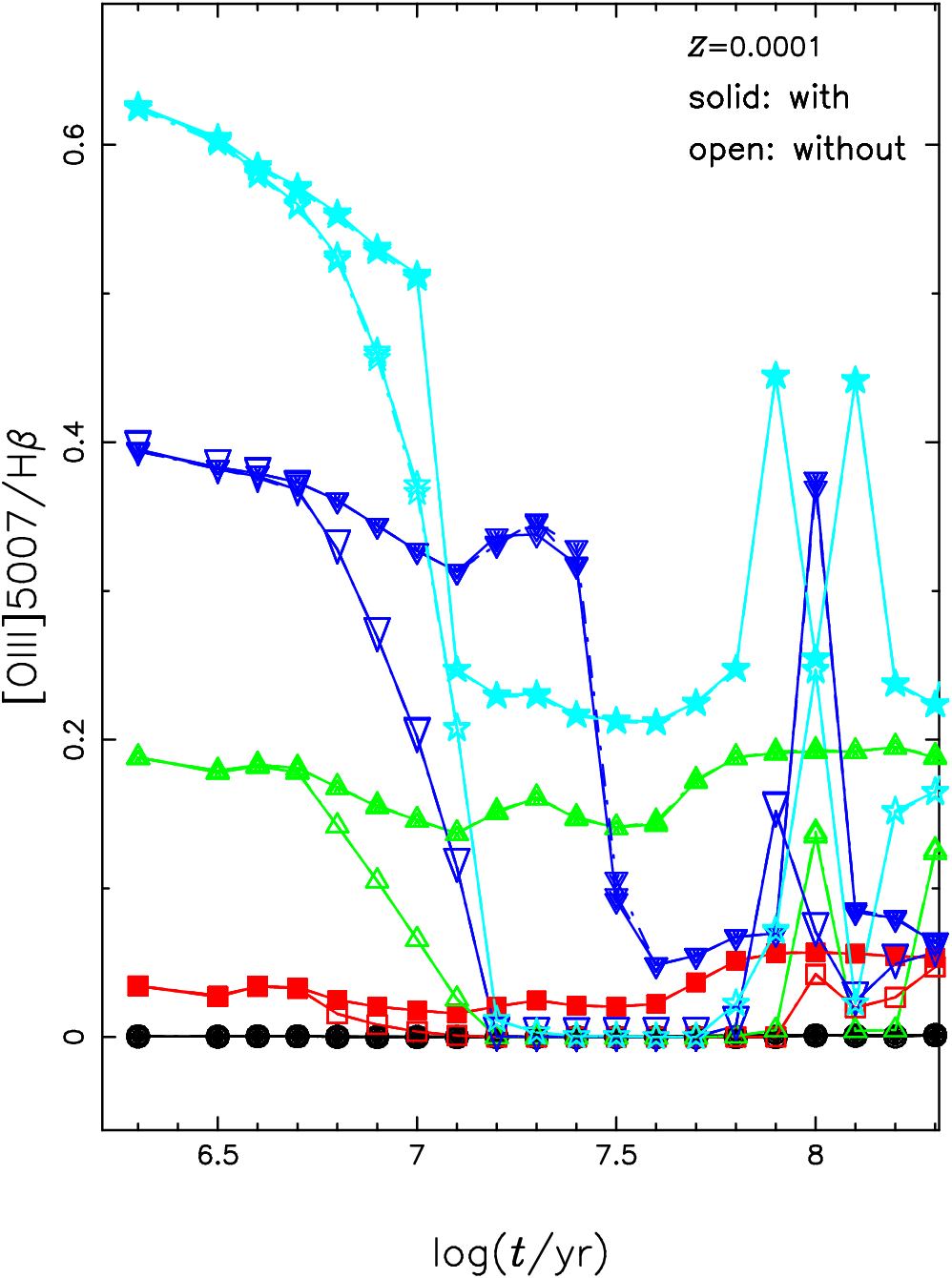} 
\includegraphics[bb=109 57 531 699,clip,height=3.8cm,width=3.6cm,angle=0]{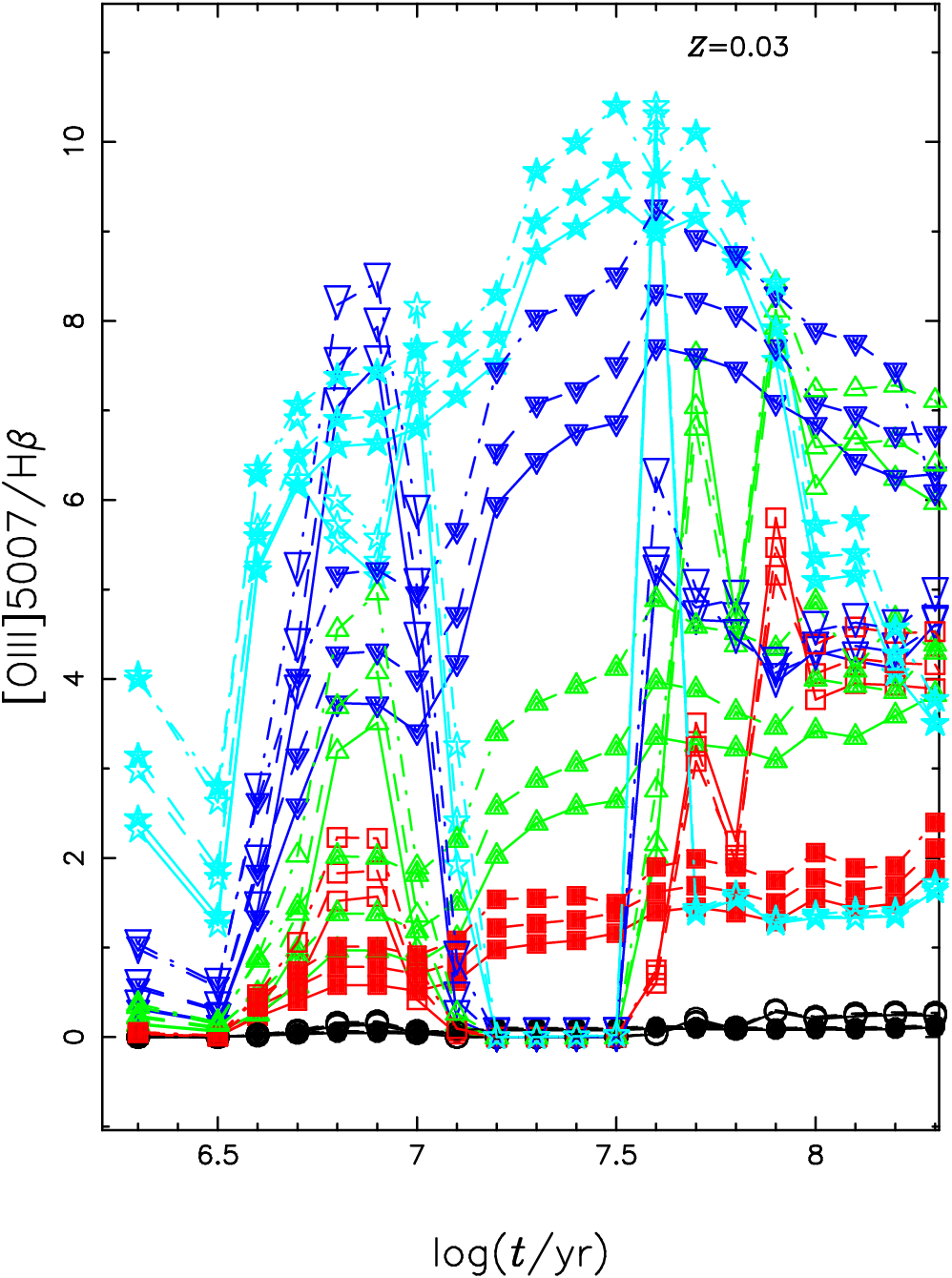}
}
\caption{
{\color{black} [O III]$\lambda$5007{\color{black}/H$\beta$} emission line strength as a function of SP's age} when using SPs with (solid symbols) and without (open symbols) BIs at $Z$=0.0001 (left panel) and 0.03 (right panel). Black [$\circ,\bullet$], red [$\Box,\blacksquare$], green [$\vartriangle,\blacktriangle$], blue [$\triangledown, \blacktriangledown$] and cyan [$\star, \bigstar$] symbols are for log$U$=$-$4, $-$3, $-$2, $-$1 and 0. Solid, dashed and dot-dashed lines are for $n_{\rm H}$=10, 100 and 350 cm$^{-3}$. }
\label{Fig:evo5007}
\end{figure}

\begin{figure}
\leftline{
\includegraphics[bb= 50  57 531 699,clip,height=3.8cm,width=4.0cm,angle=0]{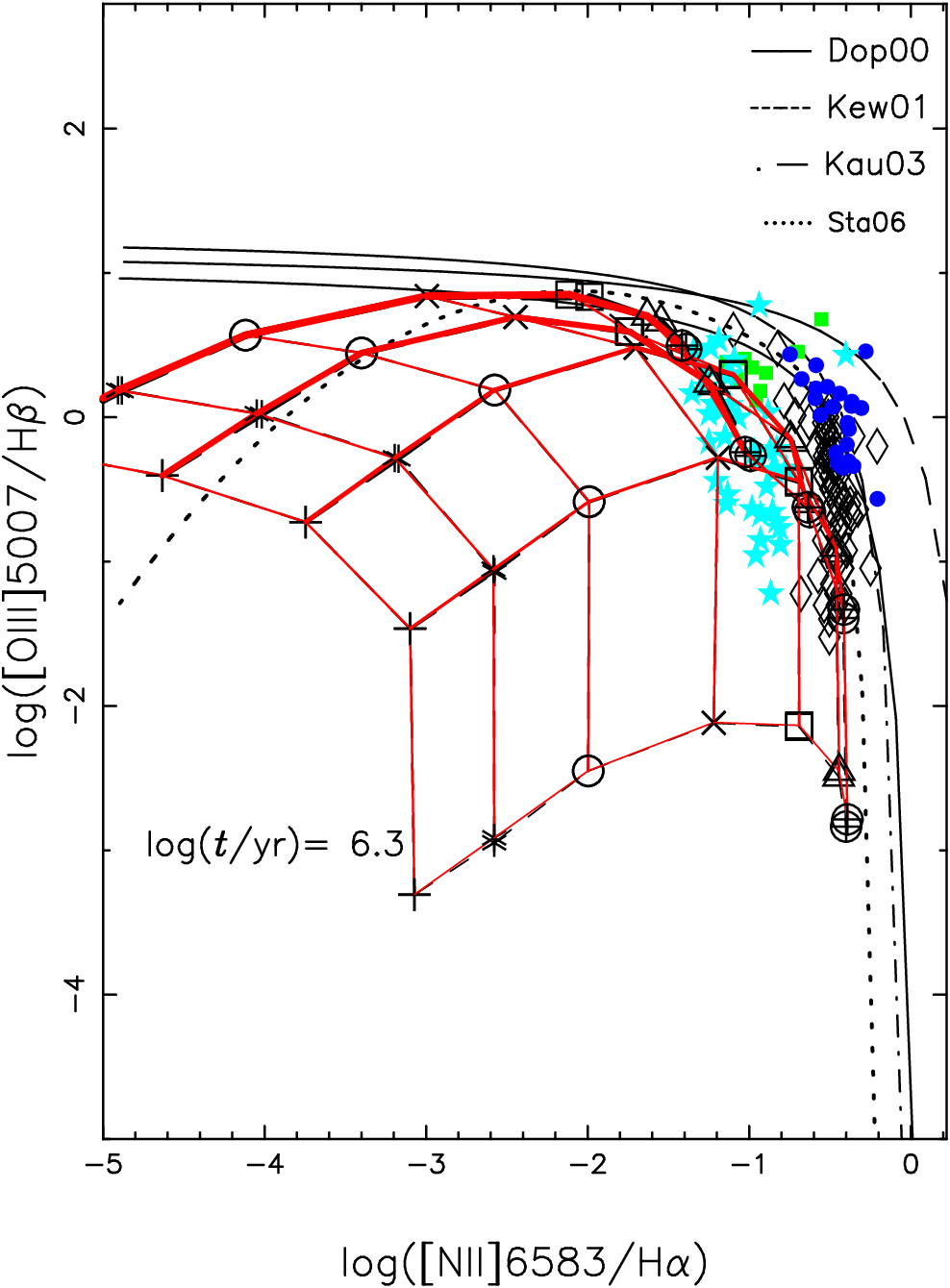} 
\includegraphics[bb=108 57 531 699,clip,height=3.8cm,width=3.6cm,angle=0]{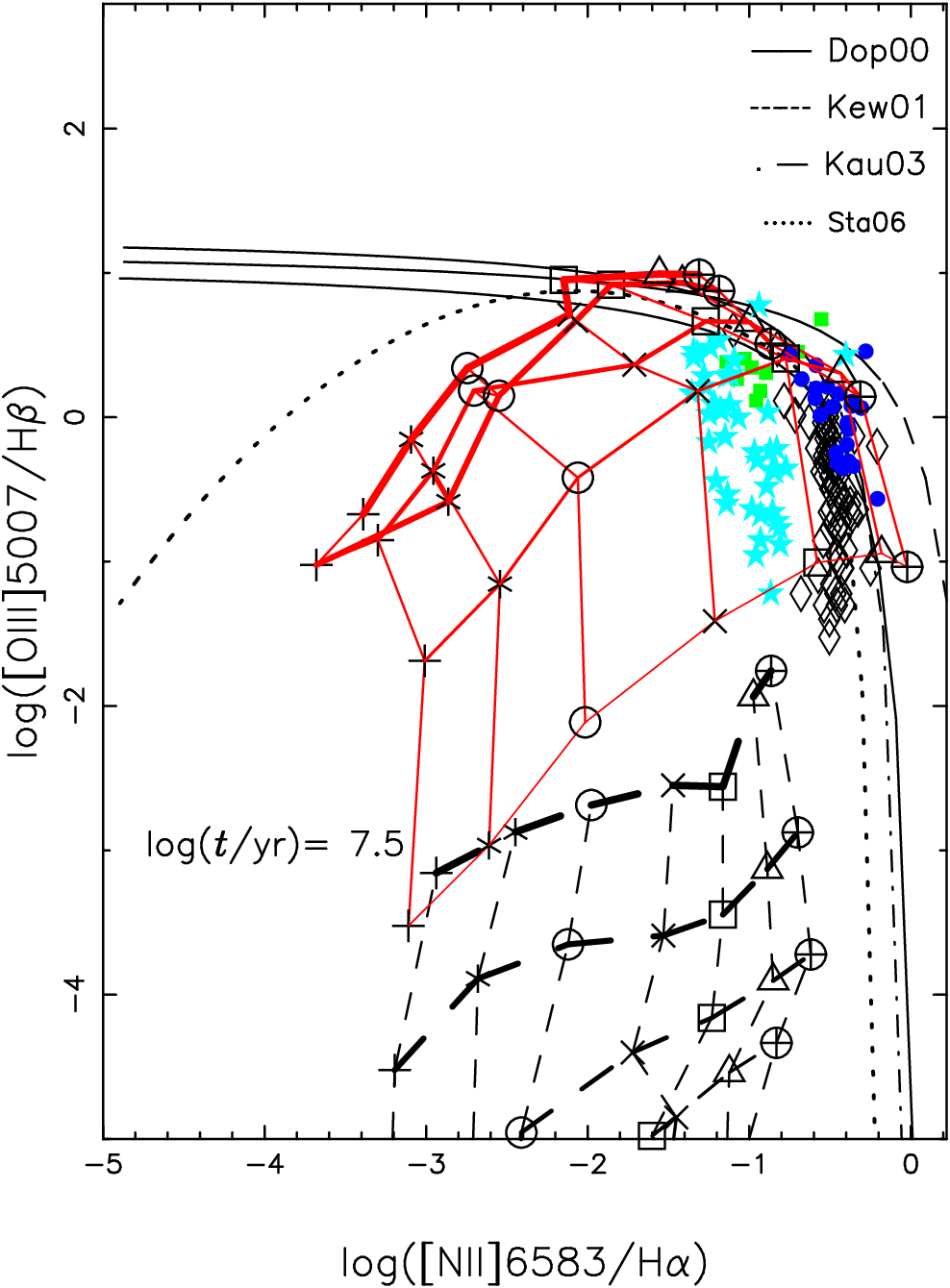}
}
\caption{VO87 diagnostic diagram log([N II]$\lambda$6583/H$\alpha$)$\sim$log([O III]$\lambda$5007/H$\beta$) when using log($t$/yr)=6.3 (young, left panel) and 7.5 (IA, right panel) SPs with (red, solid line) and without (black, dashed line) BIs in the case of $n_{\rm H}$=100${\rm cm^{-3}}$.
On each grid, symbols (+, $\ast$, $\circ$, $\times$, $\Box$, $\triangle$, $\oplus$) are for $Z$=0.0001, 0.0003, 0.001, 0.004, 0.01, 0.01 and 0.03 and line width increases with log$U$ (=$-$4,$-$3,$-$2, $-$1 and 0, from bottom to top). 
Comparisons include the border lines (Dop00, Kew01, Kau03, Sta06; --- , -\,-\,-, -$\cdot$-$\cdot$, $\cdots$; grey) and observations (B05 [$\lozenge$], V98 [{\color{black} $\blacktriangle$}], S13[{\color{green} $\blacksquare$}], S10[{\color{blue} $\bullet$}], B09[{\color{cyan} $\bigstar$}], R08[{\color{magenta} $\square$}], W11[{\color{yellow} $\circ$}], V06[{\color{red} $\bigcirc$}]). 
The absence of V89, R08, W11 and V06 observations in this plot is caused by the lack of corresponding data.}
\label{Fig:o3n2-dt}
\end{figure}

\begin{figure}
\leftline{
\includegraphics[bb= 50  57 531 699,clip,height=3.8cm,width=4.0cm,angle=0]{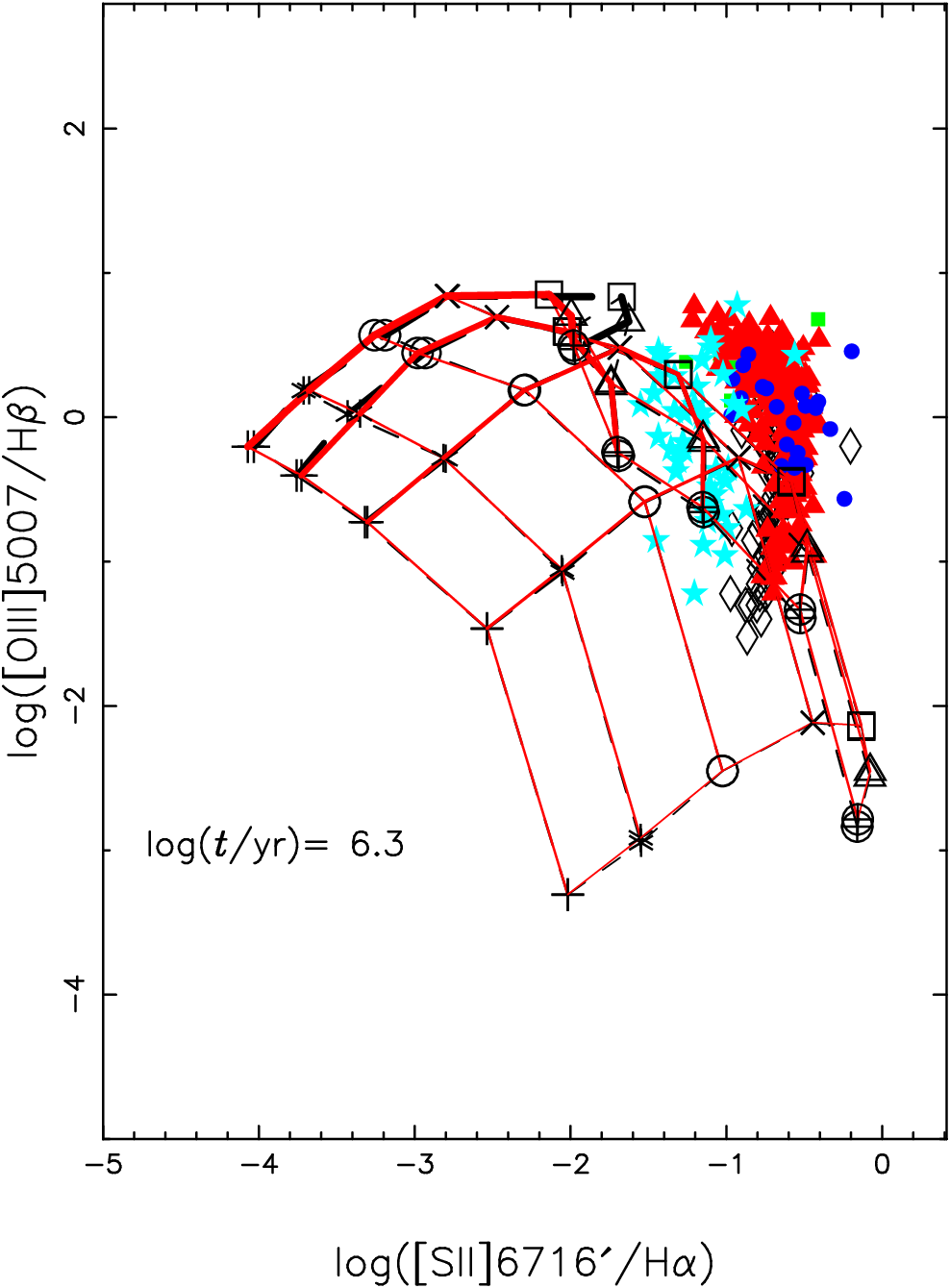} 
\includegraphics[bb=108 57 531 699,clip,height=3.8cm,width=3.6cm,angle=0]{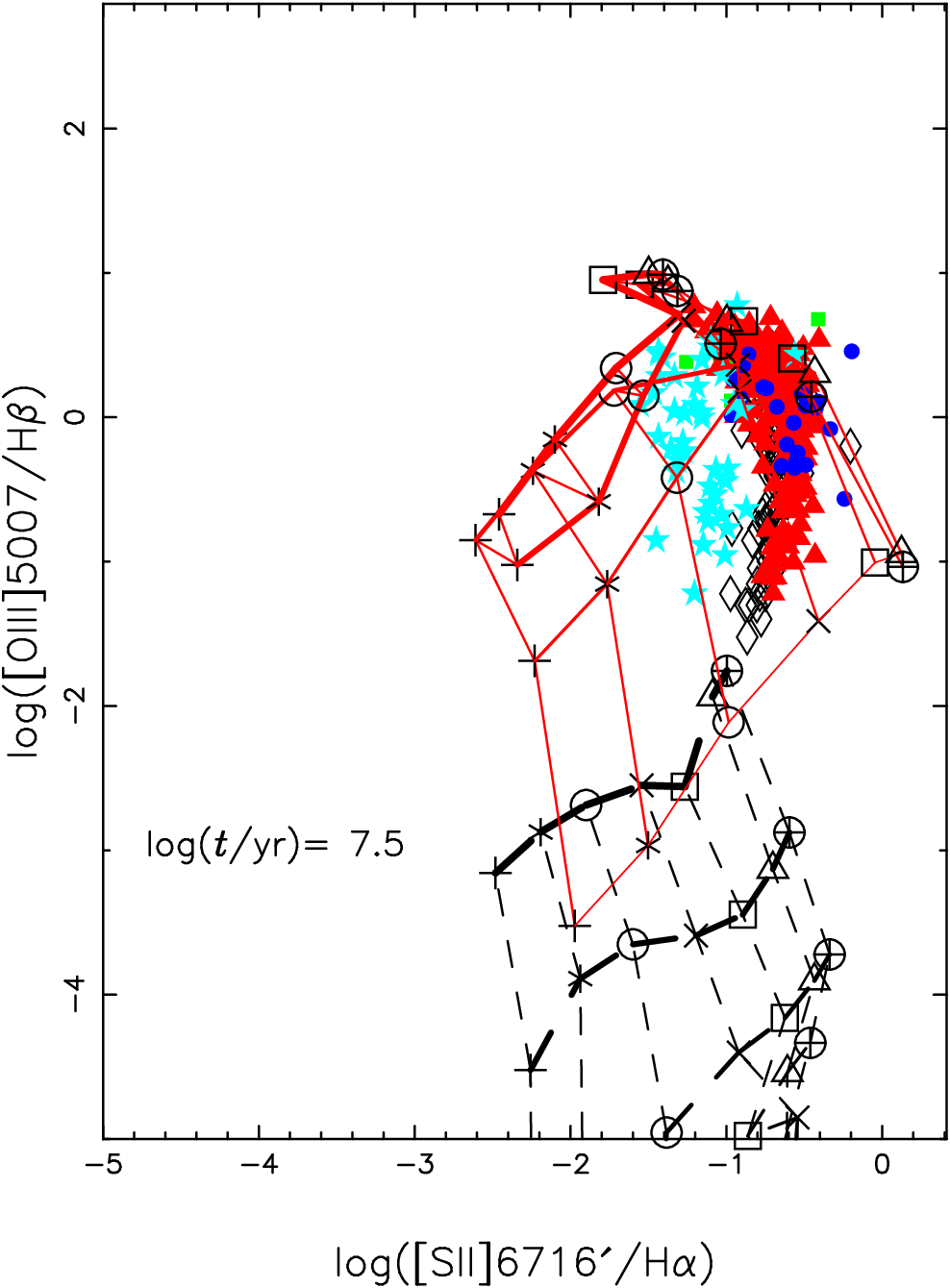}
}
\caption{Similar to Fig.~\ref{Fig:o3n2-dt}, but for log([S II]$\lambda$6716$'$/H$\alpha$)$\sim$log([O III]$\lambda$5007/H$\beta$).}
\label{Fig:o3s2-dt}
\end{figure}

In this section, we will show that {\color{black} radiation fields emitted by IA [7$\la$log($t$/yr)$\la$8] SPs with binaries {\color{black} are in theory possible candidates of} significant} CISs of classic H II regions and show the other {\color{black} resultant features}.
This work will include various metallicity ($Z$), electron-temperature ($T_{\rm e}$), electron-density ($n_{\rm e}$) and ionization-parameter ($U$) sensitive emission lines,  constituent lines, line strength ratios and grids (Table~\ref{Tab:line-list}, compiled from literatures).
In total, we calculate  8190 models (Section 2.3), including different $U$, $n_{\rm H}$, $Z$, $t$ and condition that BIs are taken into account.

The conjecture, {\color{black} radiation fields of IA SPs with BIs are possible candidates of significant} CISs of H II regions, is based on the facts that the inclusion of BIs can produce some hot [log($T_{\rm eff}$/K)$\sim$5] and luminous He MS stars at  intermediate ages (see Fig. 4 of \citealt[][hereafter {\color{magenta} Z12}]{zha12}) and that this kind of object makes the UV and EUV spectra harder. 
In Fig.~\ref{Fig:ised-bi}, we give the spectrum evolution (91$\le$$\lambda$/{\rm \AA}$\la$1100) of SPs with and without BIs at $Z$=0.02, the SP's age is within log($t$/yr)=5.1-10.18. 
From it, the following phenomena can be seen. 
%
%
(i) When log($t$/yr)$\la$$t_1$ (=6.8$\sim$7.0 for $Z$=$10^{-3}$$\sim$0.02) or $\ga$$t_2$ (=7.8$\sim$7.7 for $Z$=$10^{-3}$$\sim$0.02), the evolution of EUV spectra is similar between SPs with and without BIs. When log($t$/yr)$\la$$t_0$ (=6.6) they are soft and then become hard until $t_1$. Note that some photoionization models (such as Dop00, Kew01, etc.) used the spectra of SPs at ages of $\sim$$t_1$. When log($t$/yr)$\ga$$t_2$, the EUV spectra are hard but their strengths are far less than young SPs (similar to the conclusion made by \citealt{kan14}).
%
(ii) When $t_1$$\la$log($t$/yr)$\la$$t_2$, the EUV spectra are completely different. The spectra of SPs without BIs are soft and their strengths are low, while for those with BIs, the opposite holds and their spectrum slope approaches to that for young SPs.
{\color{black} The reason of the differences in Fig. 1 can be seen in Sections 3 (2nd paragraph) and 4. In} following analyses, we only focus on early and intermediate ages.

\subsection{\color{black} Possibility: SPs with BIs are significant CISs}
\label{Subsect:}
In this section, we will present the effect of BIs on various line strengths/strength-ratios and show {\color{black} in theory the possibility} that {\color{black} radiation fields of IA SPs with BIs are significant} CISs of  H II regions.

First, we will discuss the effect of BIs on various $U$, $Z$, $n_e$ and $T_e$ sensitive line strengths/strength-ratios.
By comparisons, we find that the differences are small between the CISs acted by {\color{black} radiation fields emitted by} young SPs with and without BIs, while large between using the IA SPs with and without BIs.
The line strengths/strength-ratios, when {\color{black} radiation fields of} IA SPs with BIs are used as the CISs, are comparable to (even greater than) those for young SPs.
This can be seen from Fig.~\ref{Fig:evo5007}, 
{\color{black} in which we give the [O III]$\lambda$5007{\color{black}/H$\beta$} (the strongest) emission line strength as a function of SP's age only at $Z=0.0001$ and 0.03 as an example for the sake of paper's size.}

Second, we will show {\color{black} in theory the possibility} that {\color{black} radiation fields of IA SPs with BIs are important} CISs of H II regions by comparing various grids between using SPs with and without BIs at early and intermediate ages.
In Fig.~\ref{Fig:o3n2-dt}, we take log([N II]$\lambda$6583/H$\alpha$)$\sim$log([O III]$\lambda$5007/H$\beta$) (i.e., diagnostic diagram of \citealt[][hereafter {\color{magenta} VO87}]{vei87}) as an example.
As comparisons, we also give the border/ selection-criterion lines between star-forming galaxies and AGNs from Dop00, Kew01, \citet[hereafter {\color{magenta} Kau03}]{kau03} and Sta06.
Also shown are the observations of \citet[V98]{van98}, \citet[B05]{bre05} for spiral, \citet[S13]{sta13} for NGC300, \citet[S10]{sta10} for M81, \citet[B09]{bre09} for M83, \citet[R08]{ros08} for M33, \citet[W11]{wer11} for H II regions and \citet[V06]{van06} for dwarf-irregular galaxies\footnote{\tiny For V98, using the line strength ratios of 4959$'$/4363 (provided), 6716/6731 (provided) and 5007/4959 (=2.88, true) and the line strengths of 4959$'$ and 6716$'$, we derive the line strengths of 4363, 4959, 5007, 6716 and 6731.}.
From them, we see that both {\color{black} radiation fields of young SPs and the IA SPs with BIs are {\color{black} in theory possible candidates of significant}} CISs of H II regions 
and that the grid derived from IA SPs without BIs can not cover the observational data region.

\subsection{Some lines and selection criteria}
\label{Sect:}
When {\color{black} radiation fields of} young SPs are used as the CISs of H II regions, the strengths of some model lines, such as [O III]$\lambda$4959$'$ and [O II]$\lambda$3727$'$ mentioned by D06, [S II]$\lambda$6716$'$ and [O III]$\lambda$5007 by Sta06, are weaker than observations. This can be seen from the left panel of Fig.~\ref{Fig:o3s2-dt}, which gives the grid of 
log([S II]$\lambda$6716$'$/H$\alpha$)$\sim$log([O III]$\lambda$5007/H$\beta$).
%
When the {\color{black} radiation fields from} IA SPs with BIs are used as the CISs, the grid partly covers the observational data region (see the right panel of Fig.~\ref{Fig:o3s2-dt}), thus this problem is partly solved.

Meanwhile, from the comparison between the left and right panels of Fig.~\ref{Fig:o3s2-dt}, we see that the grid border moves {\color{black} toward the upper-right corner} when {\color{black} radiation fields of} young SPs are replaced by {\color{black} those of} IA SPs with BIs as the CISs. This means that the criterion of selecting star-forming galaxies from AGNs  would move {\color{black} toward the upper-right corner}. This also can be seen from Fig.~\ref{Fig:o3n2-dt}.

%
\subsection{He II $\lambda$1640 line}
Broad He II $\lambda$1640 emission line is observed in Lyman break galaxies \citep{sha03} and some high-redshift gravitationally lensed galaxies {\color{black} \citep{cab08,des10}. 
He II feature can be produced by shocks, X-ray binary evolution and WR/massive stars \citep{bri08}.}
\citet{dop11} found that narrow He II feature can be produced by radiative shocks (from \citealt{kew13}). 
In this work, we find that He II $\lambda$1640 line can be produced when {\color{black} radiation fields of}  IA SPs with BIs serve as the CISs of H II regions.

\section{Analyses and Discussions}
\label{Sect:}
%

We will analyze the reasons of the above results. They are caused generally by the fact that the line strengths increase when {\color{black} radiation fields of} IA SPs with BIs are used as the CISs of H II regions.
The emission line spectrum of an H II region is determined primarily by the effective temperature of the cluster stars and by the ionization parameter \citep{dop05}.
The inclusion of BIs can produce very hot and luminous He MS stars in SPs at intermediate ages (see Fig. 4 of  Z12), causing that the SPs' spectrum hardness in UV band and the number of ionizing photons in nebulae {\color{black} $Q$(H)} are significantly increased (by $\sim$2\,dex, see Fig. 2 of Z12.
By the way, this value is significantly greater than the value of $\sim$0.5\,dex caused by metallicity, which is estimated from Fig. 2 of Z12, Fig. 3 and Table 3 of \citealt{zha13}), so more ions are ionized and the emission line strengths increase.
Meanwhile, when {\color{black} radiation fields of} IA SPs serve as the CISs, the structure of H II regions ($T_{\rm e}$, $n_{\rm e}$, radius, element abundance distributions, etc.) also changes.
%
{\color{black} Moreover, it should be noted that log$Q$(H)$\sim$46 when log($t$/yr)$\la$$t_1$ and log$Q$(H)$\sim$43.5 or 41 when log($t$/yr)$\sim$$t_2$ for solar-metallicity SPs with and without BIs,
this implies that the amount of gas ionized and the fluxes of gas emission lines (H$\alpha$, [O II] and [O III] nebular lines) would decrease by 2.5 or 4 orders of magnitude, for identical gas properties and CIS' geometry.}

%

In our calculations, blackbody spectrum is used for very hot stars in Yunnan-II EPS models.
Moreover, when using {\sl MAPPINGS} code to calculate the line strengths, we do not consider the change in the mechanical energy $L_{\rm mech}$ caused by BIs. 
If these factors are considered, the results will be slightly changed but our conclusions will not be changed.

\section{Summary and conclusions}
\label{Sect:}
In this letter, we 
{\color{black} showed that radiation fields of IA [7$\la$log($t$/yr)$\la$8] SPs with BIs are in theory possible candidates of} significant CISs of classic H II regions by using photoionization {\sl MAPPINGS} code and Yunnan-II EPS models.
Besides this, the border /selection-criterion lines, between star-forming galaxies and AGNs in the diagnostic diagram, would move into the region occupied originally by AGNs,  {\color{black} the theoretical strengths of some lines} (such as {\color{black} [O III]}$\lambda$4959$'$, [S II]$\lambda$6716$'$, etc.), weaker than observations, would be raised, and He II $\lambda$1640 line can be produced.
All data will be uploaded to our website (www1.ynao.ac.cn/$\sim$zhangfh).

\section*{acknowledgements}
This work was funded by the Chinese Natural Science Foundation (Grant Nos 11273053, 11073049, 11033008 \&11373063) and by Yunnan Foundation (Grant No 2011CI053).
{\color{black} We are also grateful to the referee for suggestions that have improved the quality of this paper.}

\small
\bibliography{zfh-mn3}

\bsp
\label{lastpage}
\end{document}